\documentclass[prl,showpacs,twocolumn]{revtex4}
\usepackage{graphicx}
\usepackage{amssymb}
\usepackage{graphicx}
\usepackage{amsmath, amssymb}
\usepackage{enumerate}
\usepackage{amsthm}
\def\>{\ensuremath{\rangle}}
\def\<{\ensuremath{\langle}}
\def\h{\ensuremath{\mathcal{H}}}
\def\r{\ensuremath{\mathcal{R}}}
\def\e{\ensuremath{\mathcal{E}}}

\def\b{\ensuremath{\mathcal{B}}}
\def\m{\ensuremath{\mathcal{M}}}

\newcommand{\spa}{{\rm span}}
\newcommand{\defeq}{\stackrel{\textrm{def}}{=}}
\newcommand{\TOT}{3\otimes3}
\newcommand{\LOCCI}{\textrm{LOCC-indistinguishable}}

\theoremstyle{plain}
\newtheorem{theorem}{Theorem}
\newtheorem{lemma}[theorem]{Lemma}

\newtheorem{proposition}[theorem]{Proposition}

\theoremstyle{definition}
\newtheorem{definition}[theorem]{Definition}

\theoremstyle{remark}

\def\squareforqed{\hbox{\rlap{$\sqcap$}$\sqcup$}}
\def\qed{\ifmmode\squareforqed\else{\unskip\nobreak\hfil
\penalty50\hskip1em\null\nobreak\hfil\squareforqed
\parfillskip=0pt\finalhyphendemerits=0\endgraf}\fi}

\newenvironment{proofof}[1]{\begin{trivlist}%
\item[]{\flushleft\em Proof of #1. }}
{\qed\end{trivlist}}

\begin{document}

\title{Characterizing locally distinguishable orthogonal product states}
\author{Yuan Feng$^{1,2}$}
\email{feng-y@tsinghua.edu.cn}
\author{Yaoyun Shi$^{2}$}
\email{shiyy@eecs.umich.edu}
\affiliation{$^1$State Key Laboratory
of Intelligent Technology and Systems, Department of Computer
Science and Technology, Tsinghua University, Beijing, 100084, China
\\
$^2$Department of Electrical Engineering and Computer Science,
University of Michigan, 2260 Hayward Street, Ann Arbor, MI
48109-2121, USA}
\date{\today}
\pacs{03.67.-a, 03.65.Ud, 03.67.Hk}

\begin{abstract}
Bennett et al.~\cite{BDF+99} identified a set of orthogonal {\em
product} states in the $3\otimes 3$ Hilbert space such that reliably
distinguishing those states requires non-local quantum operations.
While more examples have been found for this counter-intuitive
``nonlocality without entanglement'' phenomenon, a complete and
computationally verifiable characterization for all such sets of
states remains unknown. In this Letter, we give such a
characterization for the $3\otimes 3$ space.
\end{abstract}

\maketitle
A pure quantum state $|\phi\>_{AB}$ of a bipartite system $AB$
is said to be entangled if it is not a product state, i.e., it
cannot be represented as $|\alpha\>_A\otimes|\beta\>_B$,
for some state $|\alpha\>_A$ and $|\beta\>_B$ of the system
$A$ and $B$, respectively.
An entangled quantum state may generate
measurement statistics that are inherently different
from those generated by a classical process \cite{EPR, Bell}.
This feature of entanglement is referred to
as the nonlocality of quantum states.
Dual to the notion of state nonlocality is the nonlocality of
quantum operations. A natural definition of a
local quantum operation on a multi-partite quantum system
is that of {\em Local Operations and
Classical Communication (LOCC)} protocols,
in which each party may apply to his system
arbitrary quantum operations, while the inter-partite
communication must be classical.
It follows from the definition that
no LOCC protocol creates quantum entanglement.
However, the reverse is false. This surprising
fact was discovered by Bennett et al. \cite{BDF+99}
and was formulated as a problem of
reliably distinguishing quantum states.

\begin{figure} \centering
\includegraphics[width=3in]{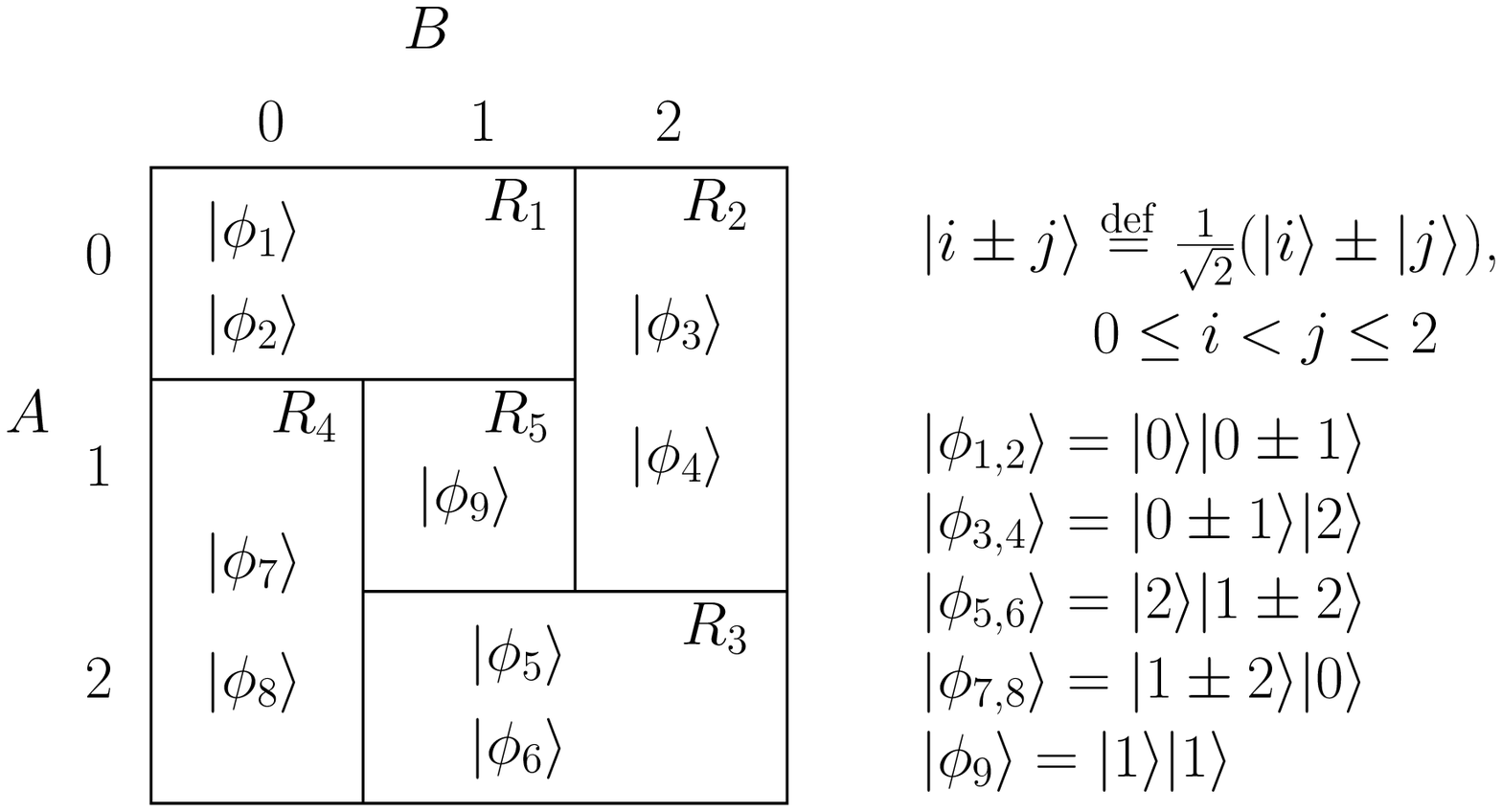}
\caption{The basis $\b_9$ for $\mathbb{C}^3\otimes\mathbb{C}^3$ and its
rectangular representation $(\r_9, \{|0\>, |1\>, |2\>\}, \{|0\>, |1\>, |2\>\}, U, V)$,
where $\r_0=\{R_i : 1\le i\le 5\}$,
$V_{R_1}$, $U_{R_2}$, $V_{R_3}$, and $U_{R_4}$ are Hadamard and the other
unitaries are Identities.
}
\label{fig:b9}
\end{figure}

A set of state $\e=\{ |\phi_i\>_{AB} \}_i$
is said to be {\em reliably distinguishable}
by a quantum operation $T$ if on each $|\phi_i\>_{AB}$,
$T$ outputs $i$ with probability $1$.
The authors of \cite{BDF+99} identified an orthonormal
basis  $\mathcal{B}_9$ for $\mathbb{C}^3\times\mathbb{C}^3$, illustrated in
Fig.~\ref{fig:b9}, that cannot be reliably distinguished by LOCC. The important feature of the basis is that each
base vector is a product state, thus the distinguishing operator cannot create entanglement.

The above property of nonlocal operations not necessarily creating
entanglement is referred to as ``nonlocality without entanglement'',
and has been studied by many authors
subsequently~\cite{BDF+99,BDM+99,DMS+03,GKR+01,WH02,HSSH03,
Ri04,Fa04,WSHV00,VSPM01,CY01,DFJ+07,Co07}. Formally, an {\em
orthogonal product set (OPS)} is a set of bipartite states which are
product states and are pairwise orthogonal. An OPS that forms a
basis is also called an {\em orthogonal product basis (OPB)}. Much
effort has been devoted to searching for additional
LOCC-indistinguishable OPSs. Besides $\mathcal{B}_9$,
Ref.~\cite{BDF+99} actually showed that
$\b_8\defeq\b_9-\{|1\>|1\>\}$ is not LOCC-distinguishable, either.
All other known $\LOCCI$ OPSs belong to the following two classes.
\begin{definition}[\cite{BDM+99}] An {\em unextendable product basis (UPB)}
is an OPS that is not a proper subset of any other OPS.
\end{definition}
Note that a UPB is not necessarily a basis for the underlying
product space. If $\e$ is an OPS in a product space
$\h_A\otimes\h_B$, denote by $\e_A = \{|\alpha\>\in\h_A :
\exists|\beta\>\in \h_B, |\alpha\>|\beta\>\in\e\}$, and
$\e_B=\{|\beta\>\in\h_B : \exists |\alpha\>\in\h_A,
|\alpha\>|\beta\> \in \e\}$.
\begin{definition}[\cite{Ri04}] An OPS $\e$ is
{\em irreducible} if neither $\e_A$ nor $\e_B$ can be partitioned into two
nonempty orthogonal subsets.
\end{definition}
\begin{theorem}[\cite{BDM+99, DMS+03, Ri04}]\label{thm:known}
The following OPSs are $\LOCCI$:
\begin{enumerate}[(1)]
\item An irreducible OPB~(\cite{Ri04}).
\item A UPB~(\cite{BDM+99, DMS+03}).
\end{enumerate}
\end{theorem}

In fact, Ref.~\cite{Ri04} characterizes all LOCC-indistinguishable OPBs.
\begin{theorem}[\cite{Ri04}]\label{thm:ri}
An OPB cannot be reliably distinguished by LOCC if and only if it
contains an irreducible subset that spans a product space.
In particular, an OPB in $3\otimes 3$ space is LOCC-indistinguishable
if and only if it is irreducible.
\end{theorem}

A main objective of this line of research is to identify
additional $\LOCCI$ OPSs. To this end,
we generalize $\b_8$ to a broader class of
$\LOCCI$ OPSs having a similar structure.
A satisfactory understanding of LOCC indistinguishability
is a complete and computationally verifiable
characterization of all such OPSs.
Clearly, any OPS in a $1\otimes n$ system, $n\ge1$,
is LOCC-distinguishable. It is also known~\cite{BDF+99} that
the same is true for any $2\otimes n$ system, $n\ge1$.
Thus $3\otimes3$ is the smallest dimension
where such a characterization is not known.
The main result of this Letter resolves this problem.
We show that when restricted to the $3\otimes3$ space,
the generalizations of $\b_8$, together with irreducible OPBs
and UPBs, are the only possible LOCC-indistinguishable OPSs.
A key step in the proof of our characterization
is to show that all irreducible OPBs
in the $3\otimes3$ space must have a representation by rectangles
similar to that of $\b_9$.

We introduce some notions for the rest of the paper.
By a slight abuse of notation, for two vectors $|\alpha\>$ and
$|\beta\>$, we write $|\alpha\>=|\beta\>$ if there exists
a non-zero $c\in\mathbb{C}$ such that $|\alpha\>=c|\beta\>$.
\begin{definition}
Two product states $|\alpha\>|\beta\>$ and
$|\alpha'\>|\beta'\>$ are said to {\em align
on the left (right)} if $|\alpha\>=|\alpha'\>$ ($|\beta\>=|\beta'\>$).
\end{definition}

Let $m, n\ge1$ be integers.
If $\e$ is an OPS in the $m\otimes n$ dimensional space
and $|\e|=mn-1$, then $\e$ can be extended to an OPB~\cite{DMS+03}.
Denote by $\e^\perp$ the unique product state that extends $\e$ to a basis.

\begin{theorem}\label{thm:8}
Let $m, n\ge1$ be integers. An OPS described below
is LOCC-indistinguishable.
\begin{enumerate}[(3)]
\item An irreducible OPS $\e$ in $\mathbb{C}^m\otimes\mathbb{C}^n$
with $|\e|=mn-1$ such that $\e^\perp$ does not align
on either side with any element in $\e$.
\end{enumerate}
\end{theorem}
\begin{proof}
Denote by $\h_A$ and $\h_B$ the state space of Alice and
and Bob, respectively. Suppose $\e=\{|\alpha_i\>|\beta_i\>:
1\le i\le mn-1\}$ and $\e^\perp=|\alpha_0\>|\beta_0\>$.
Suppose that $\e$ can be reliably distinguished by an LOCC protocol.
Fix such a protocol $\mathcal{P}$ that takes the smallest number of rounds of communication.
Without loss of generality, assume that Alice sends the first message,
which is the measurement outcome $k$ of a
Positive-Operator-Valued Measurement (POVM)
$\m\defeq\{M_k:\h_A\to\h'_A\}_k$, where $\h'_A$ is Alice's state
space after applying $\m$ and the operators
$M_k$ satisfy $\sum_{k} M_k^\dag
M_k = I_{\h_A}$. If for each $k$,
there exists $\mu_k>0$ such that
$M_k^\dag M_k= \mu_k I_{\h_A}$, then $\sum_k\mu_k=1$
and each $M_k$ is an isometric
embedding. Thus $\m$ can be implemented by having Bob send
the message instead:
he generates a random number $k$ with probability $\mu_k$,
sends it to Alice, who applies $M_k$ to $\h_A$. This contradicts
the assumption that $\mathcal{P}$ takes the smallest number of rounds.
Therefore, there exists a $k$ such that $M_k^\dag M_k$ has $k_0\ge2$
number of distinct eigenvalues.
Fix such a $k$ for the rest of the proof.

Since the post-measurement states must
remain orthogonal so that they can be reliably distinguished
by the remaining steps of $\mathcal{P}$,
we have $\<\alpha_i|\<\beta_i|(M_k^\dag
M_k\otimes I_{\h_B}) |\alpha_j\>|\beta_j\>= 0$,
for all $1\leq i<j\leq mn-1$.
Note that $\e'\defeq \e \cup\{\e^\perp\}$ is an OPB, thus
for each $i$, $1\le i\le mn-1$,
there exist $\lambda_i, \lambda_i^0\in\mathbb{C}$,
such that
$M_k^\dag M_k\otimes I_B |\alpha_i\>|\beta_i\>=\lambda_i |\alpha_i\>|\beta_i\>
+ \lambda_i^0 |\alpha_0\>|\beta_0\>$.
Applying $\< \alpha_0|$ on both sides,
we have $\<\alpha_0|M_k^\dag M_k |\alpha_i\> |\beta_i\>=\lambda_i
\<\alpha_0|\alpha_i\>|\beta_i\> + \lambda_i^0 |\beta_0\>$.
It follows that $\lambda_i^0=0$, since $|\beta_i\>\neq |\beta_0\>$.
Therefore, $\e_A$ is a set of
eigenstates of $M_k^\dag M_k$.

If $\e_A$ does not span $\h_A$, let $|\alpha\>\in\h_A$ be a state
orthogonal to $\spa(\e_A)$. Let $|\beta\>\in \h_B$ be orthogonal to
$|\beta_0\>$. Such $|\beta\>$ must exist since otherwise
$\dim(\h_B)=1$, and $\e$ would be reducible. Then
$|\alpha\>|\beta\>$ is orthogonal to $\e'$, a contradiction to $\e'$
being a basis for $\h_A\otimes \h_B$. Therefore, $\e_A$ spans $\h_A$,
and is a complete spectrum of $M_k^\dag M_k$. It follows that $\e_A$
can be partitioned into $k_0$ number of pair-wise orthogonal
subsets, each of which corresponds to a distinct eigenvalue of
$M_k^\dag M_k$. Since $k_0\ge2$, this contradicts the assumption
that $\e$ is irreducible. Therefore, $\e$ is LOCC-indistinguishable.
\end{proof}

As mentioned above, the $\TOT$ space is the smallest space having
LOCC-indistinguishable OPSs. We also know the following useful
facts.
\begin{proposition}[\cite{DMS+03}]\label{prop:le4} An OPS
$\e$ in
$\mathbb{C}^3\otimes\mathbb{C}^3$
is LOCC-distinguishable
if $|\e|\le4$.
\end{proposition}
\begin{theorem}[\cite{BDM+99, DMS+03}]
\label{thm:upb}
Any UPB in
$\mathbb{C}^3\otimes\mathbb{C}^3$
must have exactly $5$ elements.
\end{theorem}
In what follows, we completely characterize all
LOCC-indistinguishable OPSs in the $3\otimes3$ space.
\begin{theorem}[Main Theorem]\label{thm:complete}
An OPS in $\mathbb{C}^3\otimes\mathbb{C}^3$ is LOCC-indistinguishable
if and only if it belongs to one of the three classes (1), (2), and (3).
\end{theorem}

Combining the above three results, an LOCC-indistinguishable
OPS in the $\TOT$ space must have precisely $5$, $8$, or $9$ elements,
each of which corresponds to belong to the classes (2), (3) and (1), respectively.
Whether or not an OPS is irreducible can be checked from the
the pairwise inner products of the state components. The same
information can be used to determine if an OPS is an UPB in the
$3\otimes3$ space~\cite{BDM+99, DMS+03}. Therefore,
whether or not an OPS belongs to (1), (2), or (3) can be determined
computationally.

To prove Main Theorem,
we first generalize the rectangular representation for $\b_9$
and derive some useful properties of the generalization.
Let $I$ and $J$ be two sets. A subset $R\subseteq I\times J$ is a
{\em rectangle} if $R=A\times B$ for some $A\subseteq I$ and $B\subseteq
J$. If $R=A\times B$,
denote by $I(R)\defeq A$ and $J(R)\defeq B$. A {\em rectangular decomposition}
of $I\times J$ is a partition of $I\times J$ into
rectangles. Fig.~\ref{fig:b9} illustrates a rectangular decomposition
for $\{0, 1, 2\}\times\{0, 1, 2\}$. We refer to this decomposition
as $\r_9$ and use the labeling scheme in the Figure for its elements.
\begin{definition}
Let $m, n\ge1$ be integers, $I\defeq\{0, 1, \cdots, n-1\}$, and
$J\defeq\{0, 1, \cdots, m-1\}$. Let $\e$ be an OPB of a product
space $\h_A\otimes \h_B$ with $\dim(\h_A)=n$ and $\dim(\h_B)=m$. A
{\em rectangular representation} of $\e$ is a quintuple $(\r,
\alpha, \beta, U, V)$ such that:
\begin{enumerate}[(a)]
\item $\r$ is a rectangular decomposition of $I\times J$.
\item $\alpha=\{|\alpha_0\>, |\alpha_1\>, \cdots, |\alpha_{n-1}\>\}$
is an orthonormal basis for $\h_A$, and similarly,
$\beta=\{|\beta_0\>, |\beta_1\>, \cdots, |\beta_{m-1}\>\}$
is an orthonormal basis for $\h_B$.
\item $U$ assigns each $R\in\r$ a unitary operator $U_R$
on $\mathrm{span}\{ |\alpha_i\> : i\in I(R)\}$, and similarly,
$V_R$ is a unitary operator on $\mathrm{span}\{|\beta_j\> : j\in J(R)\}$.
\item $\e=\left\{(U_R|\alpha_i\>)\otimes (V_R|\beta_j\>) : R\in\r, (i, j)\in R\right\}$.
\end{enumerate}
\end{definition}

It can be verified by direct inspection from Fig.~\ref{fig:b9}
that $\b_9$ has a rectangular representation of which the
rectangular decomposition is $\r_9$ and the unitary
transformations are either Identity operators or Hadamard.
Removing any state other than $|1\>|1\>$
from $\b_9$ results in an LOCC-distinguishable set.
The same is true for any OPB having a rectangular representation
using $\r_9$.
\begin{proposition}\label{prop:dist} Let $\e$ be an OPB
in the $3\otimes 3$ space having a rectangular representation
$(\r_9, \alpha, \beta, U, V)$. Suppose $|\alpha_1\>|\beta_1\>\in\b$
is the state corresponding to the $1\times1$ rectangle. Then any OPS
obtained from $\e$ by removing some state other than
$|\alpha_1\>|\beta_1\>$ is LOCC-distinguishable.
\end{proposition}
\begin{proof} We denote the states in $\e$
by $\{|\phi_i\> : 1\le i\le 9\}$ using the labeling scheme in
Fig.~\ref{fig:b9}. Without loss of generality, assume that
$|\phi_1\>$ is the only state in $\e$ missing in $\e'$. By direct
inspection, the following LOCC protocol identifies an unknown input
state from $\e'$. Bob starts the protocol by measuring
$\left\{|\beta_0\>\<\beta_0|, \ I-|\beta_0\>\<\beta_0|\right\}$. If
the measurement outcome corresponds to the first operator, Alice
measures $\left\{|\alpha_0\>\<\alpha_0|, \
U_{R_4}|\alpha_1\>\<\alpha_1| U_{R_4}^\dagger, \ U_{R_4}
|\alpha_2\>\<\alpha_2| U_{R_4}^\dagger\right\}$, concluding that the
input state is $|\phi_2\>$, $|\phi_7\>$, or $|\phi_8\>$ accordingly.
In the other case, the protocol continues using a similar strategy.
\end{proof}

We now present our Main Lemma, which characterizes
irreducible OPBs (thus LOCC-indistinguishable OPBs)
in terms of rectangular representations.
\begin{lemma}[Main Lemma]\label{lm:rec}
Any irreducible OPB in the $3\otimes 3$ space
has a rectangular representation using $\r_9$.
\end{lemma}

\begin{proof} Let $\e=\{|\alpha_i\>|\beta_i\>:
1\le i\le 9\}$ be an irreducible OPB in the $3\otimes3$
space $\h_A\otimes \h_B$.
If $|\alpha_i\>=|\alpha_j\>$, denote the state by $|\alpha_{i, j}\>$.
We will construct a rectangular representation
$P=(\r_9, \{|0\>_A, |1\>_A, |2\>_A\}, \{|0\>_B, |1\>_B, |2\>_B\}, U, V)$ for $\e$.

We first note that there exist two states
$|\alpha_1\>|\beta_1\>$ and $|\alpha_2\>|\beta_2\>\in\e$
that are aligned in at least one side.
(In fact, we can prove that in the $\TOT$ space,
there are at most 5 orthogonal product states
such that no pair of them align on either side.)
Assume that $|\alpha_1\>=|\alpha_2\>=|\alpha_{1, 2}\rangle$;
the other case would lead to the same conclusion.
Then $|\beta_1\>\perp |\beta_2\>$.
If there are $6$ states whose component in $\h_A$
is orthogonal to $|\alpha_{1, 2}\rangle$, then
they must span $(\spa\{|\alpha_{1, 2}\>\})^\perp\otimes\h_B$,
contradicting the assumption that $\e$ is irreducible.
Thus there are $|\alpha_3\>, |\alpha_4\>\in\e_A$
with $\< \alpha_{1, 2}|\alpha_3\>\ne 0$ and
$\<\alpha_{1, 2}|\alpha_4\>\ne 0$. This implies
$|\beta_3\>=|\beta_4\>$.

Repeating the above argument, we find in $\e$ pairs
of states $\{|\alpha_{5, 6}\>|\beta_5\>, |\alpha_{5, 6}\>|\beta_6\>\}$ and
$\{|\alpha_7\>|\beta_{7, 8}\>, |\alpha_8\>|\beta_{7, 8}\>\}$.
By direct inspection, $|\alpha_i\>|\beta_i\>$, $1\le i\le 8$, must be
distinct. Denote the remaining state in $\e$ by
$|\alpha_9\>|\beta_9\>$.

Let $S_A\defeq\{|\alpha_{1, 2}\>, |\alpha_9\>, |\alpha_{5, 6}\>\}$.
We show that $S_A$ is an orthonormal basis for $\h_A$. If
$|\beta_9\>=|\beta_{3, 4}\>$, $\{|\alpha_3\>|\beta_{3, 4}\>,
|\alpha_4\>|\beta_{3, 4}\rangle, |\alpha_9\>|\beta_9\>\}$ would span
$\h_A\otimes\spa\{|\beta_{3, 4}\>\}$, contradicting $\e$ being
irreducible. Thus $|\beta_9\>\ne|\beta_{3, 4}\>$, implying that for
some $i\in\{1, 2\}$, $\<\beta_i|\beta_9\>\ne 0$. Thus
$|\alpha_9\>\perp |\alpha_{1, 2}\>$. Similarly, $|\alpha_9\>\perp
|\alpha_{5, 6}\>$. If $|\alpha_{1, 2}\rangle\not\perp|\alpha_{5,
6}\>$, $\{|\beta_i\> : i=1, 2, 5, 6\}$ would be mutually orthogonal,
contradicting $\dim(\h_B)=3$. Thus $|\alpha_{1, 2}\>\perp|\alpha_{5,
6}\>$. Therefore, $S_A$ is an orthonormal basis for $\h_A$.
Similarly, $S_B\defeq\{|\beta_{7, 8}\>, |\beta_9\>, |\beta_{3,
4}\>\}$ is orthonormal in $\h_B$. Relabel $S_A$ as $\{|i\>_A : 0\leq
i\leq 2\}$ and $S_B$ as $\{|j\>_B : 0\leq j\leq 2\}$ such that
$|0\>_A=|\alpha_{1, 2}\>$, $|0\>_B=|\beta_{7, 8}\>$, etc.

Define the following unitaries as the Identity
operator on the corresponding dimension $1$ space:
$U_{R_1}$, $V_{R_2}$, $U_{R_3}$, $V_{R_4}$, $U_{R_5}$, and $V_{R_5}$.
Define
$V_{R_1}\defeq |\beta_1\>\<0|+|\beta_2\>\<1|$,
$U_{R_2}\defeq |\alpha_3\>\<0|+|\alpha_4\>\<1|$,
$V_{R_3}\defeq |\beta_5\>\<1|+|\beta_6\>\<2|$, and
$U_{R_4}\defeq |\alpha_7\>\<1|+|\alpha_8\>\<2|$.
This completes the construction of $P$.
By direct inspection, $P$ is a rectangular representation of $\e$.
\end{proof}

We are now ready to prove Main Theorem.
\begin{proofof}{Theorem~\ref{thm:complete}}
Since the ``if'' direction is precisely the combination of
Theorems~\ref{thm:known} and \ref{thm:8}, we need only to prove the
``only if'' direction. Suppose there exists an
LOCC-indistinguishable OPS $\e$ in the $3\otimes3$ space not
belonging to any of (1), (2), and (3). Then by
Proposition~\ref{prop:le4}, Theorems~\ref{thm:ri} and \ref{thm:upb},
$5\le |\e|\le 8$ and $\e$ is extensible to an OPB $\e'$. Since $\e'$
must be LOCC-indistinguishable (and thus irreducible), it has a
rectangular representation using $\r_9$, by Lemma~\ref{lm:rec}.
Since $\e$ does not belong to Class (3),
there exists a state $|\alpha\>|\beta\>$ in $\e'-\e$
not contained in the rectangle $R_5$.
Thus $\e'-\{|\alpha\>|\beta\>\}$ is LOCC-distinguishable, by
Proposition~\ref{prop:dist}. So must be $\e$, which is a
contradiction. Thus any LOCC-indistinguishable OPS must belong to
(1), (2), or (3).
\end{proofof}

Our method can also be used to give an alternative proof for
the fact that there is no LOCC-indistinguishable OPSs in $2\otimes n$
spaces observed in Ref.~\cite{BDF+99}. It remains an open problem to extend our result to 
the complete collection of LOCC-indistinguishable OPSs in spaces
of a dimension higher than $\TOT$.
To this end, it may be difficult to
extend our technique as the rectangular representation lemma
is not true for all dimensions.
For example, for any $\theta$, $0<\theta<\pi/2$ and $\theta\neq\pi/4$,
one can show that the following OPB in the $2\otimes 4$ dimensional space
does not have a rectangular representation:
\begin{eqnarray*}\label{eq:exam2}
\begin{array}{rcl}
|\psi_{1, 2}\>&=&|0\>\otimes |0\pm1\>,\\
|\psi_{3, 4}\>&=&|1\>\otimes (\cos\theta |0\>\pm\sin\theta |1\>),\\
|\psi_{5, 6}\>&=&|0+1\>\otimes |2\pm3\>,\\
|\psi_{7, 8}\>&=&|0-1\>\otimes (\cos\theta |2\>\pm\sin\theta |3\>).\\
\end{array}
\end{eqnarray*}
One may generalize the notion of rectangular representations
through a recursive definition. Unfortunately, there also exist
OPBs that do not admit such a generalized rectangular representation.
We note that an even more general concept
is that of {\em unwindability}, defined by
DiVincenzo and Terhal~\cite{DT00}. Therefore, a deeper understanding
of unwindable OPSs may lead to a better understanding of
LOCC-indistinguishable OPSs in higher dimensions.

Our result can be interpreted as an indication that LOCC protocols
are quite powerful. Along this line, Walgate \textit{et al.}
\cite{WSHV00} proved that LOCC is sufficient to reliably distinguish
{\em two} multi-partite orthogonal pure states, even when they are
entangled. When the two states are not orthogonal, LOCC protocols
can reach the global optimality in either conclusive discrimination
\cite{VSPM01} or inconclusive but unambiguous discrimination
\cite{CY01}. Therefore, perhaps the whole class of
LOCC-indistinguishable OPSs has much simpler structure than one
may fear.

There are bipartite operators other than those distinguishing
OPSs that do not create entanglement. Thus it remains an open
problem to characterize all such operators that cannot be realized by LOCC,
even in the $3\otimes3$ dimension case.

We observe that if an OPB has a rectangular representation $(\r,
\alpha, \beta, U, V)$, then there is a simple LOCC protocol to
identify an unknown state given {\em two} copies of it: the first
copy is projected to the bases $\alpha$ and $\beta$ so that the
rectangle $R$ containing the state is identified, then the second
copy is measured in the product basis $\{U_R|\alpha_i\>\otimes
V_R|\beta_j\> : (i, j)\in R\}$. Given an OPS, determining the number
of copies of an unknown state necessary to admit an LOCC distinguishing
protocol is an interesting generalization of determining if it is
LOCC-distinguishable.

Another interesting generalization is to determine the optimal
probability of identifying an unknown state from a given OPS by
LOCC. Finally, it remains possible that an operator cannot be realized
by LOCC yet may be approximated to an arbitrary precision.
Identifying such an operator or proving that none exists
is a fascinating open problem.

We thank Runyao Duan and Zhengwei Zhou for discussions, and for
pointing out related works. Y. Shi thanks Peter Shor for hosing him
at MIT, where part of this work was done. This work was partially
supported by National Science Foundation of the United States under
Awards~0347078 and 0622033. Y. Feng was also partly supported by the
FANEDD under Grant No.~200755, the 863 Project under Grant 
No.~2006AA01Z102, and the Natural Science Foundation of China under Grant
Nos.~60621062 and 60503001.

\end{document}